\documentclass[aps,twocolumn,amsmath,amssymb,floatfix,superscriptaddress]{revtex4}

\usepackage{hyperref}
\usepackage{verbatim}
\usepackage{graphicx}
\usepackage{bm} 
\usepackage{ifpdf} 
\usepackage{dcolumn}  
\usepackage{color}

\begin{document}

\title{Fluctuation-induced interactions in nematics with disordered anchoring energy}

\author{Fahimeh Karimi Pour Haddadan}
\affiliation{Faculty of Physics, Kharazmi University, Karaj 31979-37551, Iran}

\author{Ali Naji}
\affiliation{School of Physics, Institute for Research in Fundamental Sciences (IPM), Tehran 19395-5531, Iran}

\author{Nafiseh Shirzadiani}
\affiliation{Faculty of Physics, Kharazmi University, Karaj 31979-37551, Iran}

\author{Rudolf Podgornik}
\affiliation{Department of Theoretical Physics, Jo\v{z}ef Stefan Institute, SI-1000 Ljubljana, Slovenia}
\affiliation{Department of Physics, Faculty of Mathematics and Physics, University of Ljubljana, SI-1000 Ljubljana, Slovenia}
\affiliation{Department of Physics, University of Massachusetts, Amherst, MA 01003, USA}

\begin{abstract}
We examine fluctuation-induced (pseudo-Casimir) interactions in nematic liquid-crystalline films confined between two surfaces, where one of the surfaces imposes a strong homeotropic anchoring (ensuring a uniform mean director profile), while the other one is  assumed to be a chemically disordered substrate exhibiting an annealed, random distribution of anchoring energies. We employ a saddle-point approximation to evaluate the free energy of interaction mediated between the two surfaces and investigate how the interaction force is influenced by the presence of disordered surface anchoring energy. It is shown that the disorder results in a renormalization of the effective surface anchoring parameter in a way that it leads to quantitative and qualitative changes (including a change of sign at intermediate inter-surface separations) in the pseudo-Casimir interaction force when compared with the interaction force in the absence of disorder.
\end{abstract}
\maketitle

\section{Introduction}

Though discovered in this context, long-range fluctuation-induced forces are not confined solely to the example of the electromagnetic field \cite{Casimir1,Casimir2}, or indeed other fields pertaining to the fundamental description of Nature \cite{Mostepanenko},  but have been formulated also in the context of correlated materials  \cite{Kardar_rev}. Among these the liquid crystalline (LC) order \cite{de-gennes,terentjev,lubensky} could serve as the prime example and interactions engendered by thermal fluctuations of the LC order parameter(s) in confined geometries with surface anchoring have been studied extensively in this context \cite{LC1,LC2}. Surface anchoring conditions were found to be important not only in determining the equilibrium phase behavior of confined LC films, a fundamental problem in LC  physics, but also in determining the effective pseudo-Casimir or fluctuation-induced interaction mediated between the bounding surfaces \cite{rudi1,rudi2,rudi3, rudi4,ziherl,rudijure,karimi,karimi2,haddadan}.  Their effect is strongest in uniformly ordered confined LC films, close to an ordering transition, that correspond most closely to long-range correlated media and are exemplified by LC dispersions in porous glasses, aerogels and polymer networks where they represent a direct analogue of the thermal electromagnetic Casimir interaction with which they share many of the fundamental characteristics.

Fluctuations, however, need not always be of a thermal nature. For the electromagnetic field, e.g., it has been recently realized that in addition to thermal electromagnetic field fluctuations, various types of quenched structural disorder in the electromagnetic properties of the
interacting surfaces lead to the appearance of very specific long-range disorder-generated  interactions that can
either enhance or mitigate the thermal fluctuation-induced forces of the Casimir type \cite{ali-rudi,rudiali,partial,disorder-PRL,jcp2010,pre2011,epje2012,jcp2012}. This structural disorder can
be for instance associated with a quenched surface charge and/or surface potential distribution, or indeed  with quenched
inhomogeneities of the dielectric properties on the surface or in the bulk of the interacting media \cite{dean1,dean2}. The thermal and structural components of the force in this case are in general not additive and lead to complicated interplay of repulsive and attractive interactions with long and short range components. That the quenched or annealed (or, in general, even partially annealed \cite{partial}) disorder in the material properties of media interacting via electromagnetic coupling can have such profound influence on their interaction came as a surprise and its fundamental consequences are still being investigated \cite{book,Review}.

The situation is quite similar in the LC milieu. Contact surfaces in confined LC films may in general
contain various degrees of chemical or structural inhomogeneities stemming from a dirty substrate with random surface pinning, competing with LC ordering and leading to situations where the LC order is subject to a random or heterogeneous boundary condition \cite{LC3,radzihovsky1,radzihovsky2}. Depending on the sample conditions, the disorder may be present in different components of the anchoring energy assumed to be of the general Rapini-Papoular type, i.e., the energy cost due to the anchoring of the nematic director ${\mathbf n}$ at a bounding surface may be given by
\begin{equation}
H_s=-{1\over 2}\int_{S} d{\mathbf x} ~W({\mathbf x})\,({{\mathbf
n}}\cdot{\mathbf e})^2  \nonumber
\end{equation}
where ${\mathbf x}$ denotes the lateral coordinates along the bounding surface of the substrate, $S$, and the so-called easy direction, corresponding to the preferred molecular alignment at the surface, is assumed to be ${\mathbf e}={\mathbf z}$, in the model considered in this paper. Structural inhomogeneities may modify locally all the parameters entering this anchoring energy: the strength of the surface anchoring energy  $W({\mathbf x})$,  the preferred anchoring axis (easy direction)  ${\mathbf e}({\mathbf x})$ \cite{edisorder}, and/or the (rough) geometry of the surface $S$ in contact with the LC film~\cite{LC3}. These inhomogeneities may be annealed,  partially annealed or quenched, depending on whether the degrees of freedom associated with them are fully thermalized with the medium (director field) or not. Such ``randomness" effects can  influence the fluctuational behavior of the director field and thus, in turn, give rise to a modified fluctuation-induced force. While this is of course a completely valid line of inquiry per se, we want to put it specifically into the perspective of the disorder-generated pseudo-Casimir interaction for electromagnetically coupled media. The choice of our model will be to some extent biased by this point of view but nevertheless remains firmly realistic in its basic assumptions.

In a previous publication \cite{edisorder}, we analyzed the effects of both quenched and annealed disorder in the distribution of the easy direction   on the fluctuation-induced interaction mediated by a nematic film between  two plane-parallel surfaces, one of which imposes a strong homeotropic anchoring (in order to ensure a uniform mean director field across the LC film), whereas the other surface imposes a random distribution of the easy direction. It was shown that in the quenched case, the disorder effects appear additively in the total interaction, while in the annealed case, the disorder effects turn out to be non-additive in the total inter-surface interaction.
Although this is in general similar to the behavior found in the context of Coulomb systems under imposed external charge disorder, we showed that the effects of the easy direction disorder are characteristically different from the charge disorder effects: while the charge disorder is dominant only at large separations \cite{disorder-PRL,jcp2010,pre2011,epje2012,jcp2012}, the easy direction disorder prevails  at intermediate inter-surface separations, leading to a more repulsive interaction force as compared with the (disorder-free) pseudo-Casimir force.

\begin{figure}
\begin{center}
\includegraphics[width=8.cm,angle=0]{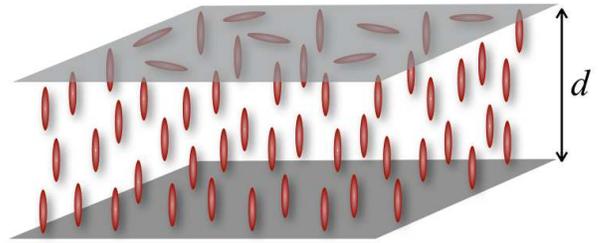}
\caption{Schematic representation of a nematic LC film in a planar cell geometry. The top surface is characterized by a disordered anchoring energy and thus the local anchoring on this plate may be homeotropic or planar with a random strength. On the bottom surface, we impose an infinitely strong homeotropic anchoring in order to ensure a uniform mean director profile.}
\label{fig1}
\end{center}
\end{figure}

In what follows, we shall focus on a model in which the disorder is present in the anchoring energy, which might be locally homeotropic or planar depending on the sign of the anchoring strength $W({\mathbf x})$ (see Fig. \ref{fig1}). Furthermore, the disorder in $W({\mathbf x})$ is considered to be annealed (the case of quenched anchoring will be studied elsewhere \cite{LC-disorder3}).
While we assume a specific type of surface disorder we nevertheless confine ourselves to the case where the easy direction remains on average homeotropic even on the ``disordered" surface at all separations between the two bounding surfaces of the cell.

We show that the disorder in the anchoring energy can modify the nature of the nematic fluctuation-induced interaction force in a way that can be described by a renormalized mean anchoring energy. The latter is determined by employing a saddle-point approximation and shows that the interaction force can differ drastically from that given by the standard, disorder-free pseudo-Casimir result. In particular, the disorder can lead to a change of sign in the interaction force in the intermediate regime of inter-surface separations as the disorder variance is increased.

The organization of the paper is as follows: In Sections \ref{sec:model} and \ref{sec:formalism}, we introduce our model and discuss the functional-integral formalism within which the interaction free energy of the system is evaluated. We proceed by studying the effects of disorder on the inter-surface force in Sections \ref{sec:force} and \ref{sec:results} and conclude our discussion in Section \ref{sec:con}.

\section{The model}
\label{sec:model}

Our model consists of a nematic LC film in a cell geometry, where the nematic phase is confined between two flat, plane-parallel surfaces located at the positions $z=0$ and $z=d$ along the normal axis to the surfaces (Fig. \ref{fig1}). The elastic energy of the LC phase is described within Frank's continuum theory in terms of a unit (headless) director field $\mathbf n({\mathbf r})$, where ${\mathbf r}=({\mathbf x}, z)$ and ${\mathbf x}=(x,y)$ are the lateral Cartesian coordinates  \cite{rudi1,rudi2,rudi3,rudi4}. We assume small fluctuations $\delta {\mathbf n}({\mathbf r})$ around the mean-field value of the director ${\mathbf n}_0({\mathbf r})$ (which is given by ${\mathbf n}_0({\mathbf r})=\hat {\mathbf z}$ in the present geometry and with a mean homeotropic anchoring on both surfaces), and hence ${\mathbf n}({\mathbf r}) = {\mathbf n}_0({\mathbf r}) + \delta {\mathbf n}({\mathbf r})$. The full Frank mesoscopic Hamiltonian can be expanded to the second order in powers of $\delta {\mathbf n}({\mathbf r})$, which, within the one-constant approximation, leads to an elastic energy expression in terms of two independent fields, $\delta n_x({\mathbf r})$ and $\delta n_y({\mathbf r})$,  that correspond to two massless (Goldstone) modes resulting from spontaneous breaking of two continuous rotational symmetry in nematics \cite{de-gennes}. For the sake of simplicity, we denote these modes by $n_i$ (for $i=1, 2$) and thus their contribution to the mesoscopic Hamiltonian can be written as
\begin{equation}
H_b={K\over 2}\sum_{i=1,2}\int_V d{\mathbf r}\, [\nabla n_i({\mathbf x},z)]^2,
\end{equation}
where the integral runs over the volume of the cell, $V$, and $K$ is the effective elastic constant. On the $z=0$ substrate, we consider a strong homeotropic anchoring, i.e., a perpendicular easy direction  ${\mathbf e}={\mathbf z}$ with infinite anchoring energy per unit area, or equivalently, the boundary condition $n_i({\mathbf x}, 0)=0$. The random anchoring energy  per unit area, $W({\mathbf x})$, is assumed to be present on the $z=d$ substrate with given statistical properties as will be specified later in this section.

The energy cost due to the surface anchoring at $z=d$ is then assumed to be given by the  Rapini-Popular surface energy, which by assuming the easy direction ${\mathbf e}={\mathbf z}$, can be written in the form of an integral over the bounding surface of the substrate, $S$, as
\begin{equation}
H_s={1\over 2}\sum_{i=1,2}\int_S d{\mathbf x}\, W({\mathbf x})[n_i({\mathbf x},d)]^2.
\label{eq:H_s}
\end{equation}
With these provisos the full Hamiltonian is  given by
\begin{equation}
H[n_1,n_2; W]=H_b[n_1,n_2]+H_s[n_1,n_2; W],
\end{equation}
and the partition function can be written in the standard functional-integral form as
\begin{equation}
{\mathcal Z}[W]=\int\bigg(\prod_{i=1,2}{\mathcal D}n_i\bigg)\,  \exp(-\beta H[\{n_i\}; W]),
\end{equation}
where  $\beta=1/(k_BT)$, $k_B$ is the Boltzmann constant, and $T$ is
the temperature.

The anchoring energy  is a random quantity and it is  assumed to be given by a Gaussian probability distribution function as
\begin{equation}
{\mathcal P}[W]=C\, e^{-{1\over 2g}\int_S d{\mathbf x}\, (W({\mathbf x})-W_0)^2},
\end{equation}
where $C$ is the normalization constant, $W_0$ (with the dimension of $[{\mathrm{energy}}]/[{\mathrm{length}}]^2$) is the mean value and $g$ (with the dimension of $[{\mathrm{energy}}]^2/[{\mathrm{length}}]^2$) is the variance of the anchoring energy per unit area  at $z=d$. Clearly, the above probability distribution function implies a spatially uncorrelated disorder. Note that $W({\mathbf x})$ on the heterogeneous surface can take both positive and negative values favoring local homeotropic or planar anchoring orientations, respectively (Fig. \ref{fig1}). On average, however, we assume a positive mean anchoring energy $W_0>0$ to prevent distortions in the mean-field profile of the director field \cite{ziherl}. We have also taken ${\mathbf e}={\mathbf z}$ on both surfaces including the disordered one (see Eq. (\ref{eq:H_s})); assuming other components for ${\mathbf e}$
makes the analysis of the fluctuation-induced force more complicated in this context as we have discussed elsewhere \cite{edisorder}. 

\section{The Formalism}
\label{sec:formalism}

In order to evaluate the free energy of the system, $F$, in the presence of {\em annealed} disorder, one has to first average
the partition function over different realizations of the disorder field, i.e., $\langle{\mathcal Z}[W]\rangle$. One then has 
\begin{equation}
F=-k_{B}T\ln \langle{\mathcal Z}[W]\rangle.
\end{equation}
 After taking the Gaussian integral over the disorder, the ``averaged" partition function $\langle {\mathcal Z}\rangle$, is obtained as
\begin{equation}
\langle {\mathcal Z}\rangle=\int\bigg(\prod_{i}{\mathcal D}n_i\bigg)\,  \exp(-\beta H_{eff}[\{n_i\}]),
\end{equation}
where the ``effective" Hamiltonian reads
\begin{eqnarray}
&& H_{eff}[\{n_i\}]={ K\over 2}\sum_{i}\int_V d{\mathbf r} \, (\nabla n_i)^2 \\
&&-\frac{1}{2}\sum_{i,j}\int_S \!d{\mathbf x} \left({\beta g\over 4} n_i^2({\mathbf x},d)n_j^2({\mathbf x},d) - W_{0}\delta_{ij}n_i^2({\mathbf x},d)\right), \nonumber
\label{bgefwyk}
\end{eqnarray}
with $\delta_{ij}$ being the Kronecker delta and $i,j=1,2$. 
As one can note, eliminating the disorder field results in a nonlinear term in the effective Hamiltonian, which is proportional to the disorder variance, $g$.  This nonlinearity can be treated in general by introducing an auxiliary  field $\lambda_{ij}({\mathbf x})$ that helps to cast  the Hamiltonian back into a quadratic form in terms of the fluctuating fields $n_i({\mathbf x},d)$~\cite{petridis,lubensky}.
We assume that the auxiliary  field has a general form as 
$\lambda_{ij}({\mathbf x}) = \lambda_{a}({\mathbf x})\delta_{ij}+\lambda_{b}({\mathbf x}) I_{ij}$, where $I_{ij}$ is a matrix with all elements equal to one. In addition and in order to make the forthcoming analytical calculations tractable, we take recourse to a mean-field approximation of the Edwards-Gupta type \cite{EdwardsGupta} by furthermore assuming that $\lambda_{a(b)}({\mathbf x}) = \lambda_{a(b)}$, where $\lambda_{a(b)}$ has no spatial dependence anymore.
After Fourier-transforming the fluctuating fields with respect to the lateral (in-plane) coordinates in a discrete description as
$n_i({\mathbf x},z)=\frac{1}{\sqrt{N}}\sum_{\mathbf q}n_{i,{\mathbf q}}(z)e^{-i{\mathbf q}.{\mathbf x}}$,
where $N=A/a^2$ is the number of lattice points of spacing $a$ over the substrate area $A$, we obtain
\begin{widetext}
\begin{equation}
\langle{\mathcal Z}\rangle=
 \bigg(\prod _{i,j}\int d\lambda_{ij}\bigg)  \bigg(\prod_{i,{\mathbf q}}\int dn_{i,{\mathbf q}}(d)\bigg) 
 \bigg(\prod_{i,{\mathbf q}}\int_{(n_{i,{\mathbf
q}}(0)=0,0)}^{(n_{i,{\mathbf q}}(d),d)}{\mathcal D}n_{i,{\mathbf q}}(z)\bigg)\,\exp\Big\{- \beta
H_{eff}\big[\{n_{i,{\mathbf q}}(z)\}, \{n_{i,{\mathbf q}}(d)\}, \{\lambda_{ij}\}\big]\Big\},
\label{eq:Z_eff}
\end{equation}
\end{widetext}
with the effective Hamiltonian  
\begin{eqnarray}
&&\beta H_{eff}\big[\{n_{i,{\mathbf q}}(z)\}, \{n_{i,{\mathbf q}}(d)\}, \{\lambda_{ij}\}\big]=\sum_{i,j}\Big\{{N\lambda_{ij}^2\over \Gamma}+\nonumber\\
&&+ \sum_{\mathbf q}\Big[{{\mathcal K}\over 2}\int_{0}^{d}dz\Big(|\partial_{z}n_{i,{\mathbf q}}(z)|^2+q^2|n_{i,{\mathbf q}}(z)|^2\Big)\delta_{ij}\
\nonumber\\
&&+n_{i,{\mathbf q}}(d)\Big({{\mathcal W}_0\over 2}\delta_{ij}-{\lambda}_{ij}\Big)n_{j,{\mathbf q}}^*(d)\Big]\Big\}.
\label{hamiltonian}
\end{eqnarray}
Here we have defined the rescaled parameters as $\Gamma=\beta^2ga^2/2$ and
\begin{equation}
{\mathcal K}=\beta Ka^2, \qquad {\mathcal W}_0=\beta W_{0} a^2.
\end{equation}
Note that ${\mathcal K}$ has the dimension of length, while ${\mathcal W}_0$ and $\Gamma$ are dimensionless.

Now since the fields appear in quadratic forms in the effective Hamiltonian, we can perform the functional integrals
over $n_i$ in Eq. (\ref{eq:Z_eff})
using the standard methods \cite{rudi1,rudi2,rudi3,rudi4,karimi,ziherl,haddadan}, which gives
\begin{equation}
\langle{\mathcal Z}\rangle=\int d\lambda_a \,d\lambda_b\,
e^{-\frac{2N}{\Gamma}[(\lambda_a+\lambda_b)^2+\lambda_b^{2}]-\frac{1}{2} \sum_{\mathbf q}{\mathrm{Tr}} \ln G_{ij}^{-1}(q)}, 
\end{equation}
where, for each mode ${\mathbf q}$, we have 
\begin{eqnarray}
&&\ln G^{-1}_{ij}(q)=\ln\left({\mathcal K}q\cosh(qd)+({\mathcal W}_0-2\lambda_{a})\sinh(qd)\right)\delta_{ij}\nonumber\\
&&+{1\over 2}\ln\left(1-{4\lambda_{b}\sinh(q d)\over {\mathcal K}q\cosh(qd)+({\mathcal W}_0-2\lambda_{a})\sinh(qd)} \right)I_{ij}.\nonumber\\
\end{eqnarray}
Hence, the partition is obtained as 
\begin{equation}
\langle{\mathcal Z}\rangle=\int d\lambda_a \,d\lambda_b\, e^{-\beta {\mathcal F}(\lambda_a, \lambda_b)}, 
\end{equation}
where we have defined 
\begin{eqnarray}
&&\beta {\mathcal F}(\lambda_a, \lambda_b)=\frac{2N}{\Gamma}\big[(\lambda_a+\lambda_b)^2+\lambda_b^{2}\big]\\
&&+ \frac{1}{2}\sum_{\mathbf q}\ln\big({\mathcal K}q\cosh(qd)+\left({{\mathcal W}_0}-2\lambda_a\right)\sinh(qd)\big)\nonumber\\
&&+ \frac{1}{2}\sum_{\mathbf q}\ln\big({\mathcal K}q\cosh(qd)+\left({{\mathcal W}_0}-2\lambda_a-4\lambda_b\right)\sinh(qd)\big).\nonumber
\end{eqnarray}

In order to evaluate the free energy of the system, $F(d)=-k_{B}T\, \ln\langle{\mathcal Z}\rangle$, we use the saddle-point approximation and minimize ${\mathcal F}(\lambda_a, \lambda_b)$ with respect to $\lambda_a$ and $\lambda_b$ while all other parameters are kept fixed \cite{petridis,lubensky}; this procedure is consistent with the mean-field assumption made for the variables $\lambda_a$ and $\lambda_b$
in the preceding steps. The minimization gives the saddle point solutions $\lambda_b^*=0$ and $\lambda_a^*=\lambda_*$, where 
$\lambda_*$ satisfies  the saddle-point equation
\begin{equation}
{2N\lambda_*\over \Gamma} -  \sum_{\mathbf q} \bigg[{{\mathcal K}q\coth(qd)}+\big({{\mathcal W}_0}-2\lambda_*\big)\bigg]^{-1} =0.
\label{lambda*1}
\end{equation}
The saddle-point free energy $F(d) =  {\mathcal F}(\lambda_*, 0)$ thus follows as
\begin{equation}
\label{eq:free_energy_2}
\beta F(d)\!\simeq \!{2N\lambda_*^2\over \Gamma}
+\! \sum_{\mathbf q}\!\ln\!\left({\mathcal K}q\cosh(qd)+\left({{\mathcal W}_0}-2\lambda_*\right)\sinh(qd)\right).
\end{equation}
Note that the disorder-free case is recovered when $g=0$ (or $\Gamma=0$), giving the solution $\lambda_*=0$.

\section{The interaction force}
\label{sec:force}

Using the above equations, the free energy can now be cast into an alternative, symmetric form which is useful in deriving the interaction force expression later, i.e.,
\begin{equation}
\beta F(d)\! =\!  \beta F_{\lambda_*}(d) - {\lambda_*\over 2}\!\left(\!\frac{\partial \beta F_{\lambda_*}(d)}{\partial {\lambda_*}}\!\right)_d,
\label{svrjkh}
\end{equation}
where we have defined $\beta F_{\lambda_*}(d) \equiv \sum_{\mathbf q}\ln({\mathcal K}q\cosh(qd)+\left({{\mathcal W}_0}-2\lambda_* \right)\sinh(qd))$.
Therefore, calculating only {\em one} quantity, $F_{\lambda_*}(d)$ , would be sufficient in order to evaluate the complete  free energy of the system. This latter quantity can be decomposed as 
\begin{equation}
F_{\lambda_*}(d) = C+I({\lambda_*})+F_C({\lambda_*}, d),
\end{equation}
where $C=\sum_{\mathbf q} \ln ({\mathcal K}q/2) + d\sum_{\mathbf q} q$ is the reference surface and bulk energy of the LC phase (which is irrelevant here), $I({\lambda_*})= \sum_{\mathbf q}\ln(1+({{\mathcal W}_0}-2\lambda_*)/{\mathcal K}q)$ is  the surface energy of the disordered interface, and
\begin{equation}
\beta F_{C}({\lambda_*}, d) = \sum_{\mathbf q}\ln\Big(1+\frac{{\mathcal K}q-({{\mathcal W}_0}-2\lambda_*)}{{\mathcal K}q+({{\mathcal W}_0}-2\lambda_*)}\,e^{-2qd}\Big)
\label{eq:F_C}
\end{equation}
is  analogous to the  pseudo-Casimir interaction free energy of a LC cell in the absence of disorder (which follows simply by setting $\lambda_*=0$ in the above equation) but with a {\em renormalized anchoring parameter} due to the presence of disorder at the surface  $z=d$, i.e.,
\begin{equation}
{\mathcal W}_{eff}\equiv {{\mathcal W}_0-2\lambda_*}.
\label{eq:eff_W}
\end{equation}
Equivalently one could thus say that the disorder renormalizes the {\em extrapolation length} to an effective value given by $1/\ell_{eff} \equiv 1/\ell - 2\lambda_* /{\mathcal K}$,  where $\ell={K}/{W_0}$ 
is the {\em mean extrapolation length} determined by the mean anchoring energy per unit area, $W_0$.
 
 In addition, we note that the above expression for $\beta F_{C}({\lambda_*}, d)$ also resembles the thermal Casimir-van der Waals  interaction free energy between a dielectric interface with a two-dimensional dipolar layer and a conductor surface placed a distance $d$ away \cite{rudi5}. In fact, in this analogy, ${\mathcal K}$ could be identified with the surface polarizability and ${\mathcal W}_0$ with the dielectric discontinuity.

The net interaction force between an annealed disordered surface and an ordered one can then be obtained from $f(d)=-\partial F(d)/\partial d$ and by making use of Eqs. (\ref{eq:free_energy_2})-(\ref{eq:F_C}) and also noting that $F(d)$ depends on $\lambda_*=\lambda_*(d)$, which is  to be differentiated properly with respect to $d$ as well. It turns out, however, that the final expression for the force can be deduced only from $F_{C}({\lambda_*}, d)$ when $\lambda_*$ is kept fixed, that is as
\begin{equation}
 \beta f(d) =
 - \left(\frac{\partial  \beta F_{C}({\lambda_*}, d)}{\partial d}\right)_{\lambda_*}\!\!\! = \sum_{\mathbf q} \frac{2q}{1+\frac{{\mathcal K}q+({{\mathcal W}_0}-2\lambda_* )}{{\mathcal K}q-({{\mathcal W}_0}-2\lambda_* )}\,e^{2qd}}.
\label{eq:force_tot}
\end{equation}
In other words, the surface energy of the disordered interface, $I({\lambda_*})$,  does not enter the final force expression as one could have expected. 

In order to proceed,  we express the above formulas in a  dimensionless continuum representation using $ \sum_{\mathbf q}\rightarrow A\int d{\mathbf q}/(2\pi)^2$
and the  following definitions for the rescaled parameters
\begin{equation}
  p = q\ell,\quad \tilde d = \frac{d}{\ell}, \quad  \Lambda_* = \frac{2\lambda_*}{{\mathcal W}_0}, \quad \chi = {g\over 4\pi K^2}.
  \label{eq:defs}
\end{equation}
Equation (\ref{lambda*1}) can then be written as
\begin{equation}
\Lambda_*=
\chi \int_{0}^{p_{max}}{p dp\over p\coth(p{\tilde d})+1
 {-\Lambda_*}},
\label{eq:big_lambda}
\end{equation}
where we have inserted the upper (ultra-violet) cut-off $p_{max}=\pi\ell/a=\pi/\tilde a$.
The interaction force per unit area (pressure) can be made dimensionless as
\begin{equation}
\Pi(\tilde d) \equiv \frac{\ell^3 \beta f(d)}{A} =
\frac{1}{\pi}\!\int_{0}^{\infty}\!\frac{p^2~dp}{1+\frac{p+(1-\Lambda_*)}{p-(1-\Lambda_*)}\, e^{2p{\tilde d}}}.
\label{eq:force_res0}
\end{equation}
It is worth noting that in this new representation, the renormalized anchoring parameter (for the the surface  $z=d$) can be written as
\begin{equation}
\tilde {\mathcal W}_{eff}\equiv \frac{{\mathcal W}_{eff}}{{\mathcal W}_0}= {1-\Lambda_*},
\label{eq:eff_W2}
\end{equation}
which enters directly into expression (\ref{eq:force_res0}) for the rescaled pressure.

\section{Results}
\label{sec:results}

In order to evaluate the interaction force (\ref{eq:force_res0}), one first needs to solve Eq.  (\ref{eq:big_lambda}) for the parameter $\Lambda_*$ as we shall do numerically later. But before proceeding further we consider the behavior of $\Lambda_*$ in the limit of large and small separations.

For large inter-surface separations, $\tilde d\gg 1$,  the saddle-point equation for $\Lambda_*$, Eq.  (\ref{eq:big_lambda}), can be solved by approximating $\coth (qd)\simeq 1$ giving
\begin{equation}
   \Lambda_* \simeq \chi p_{max} - \chi(1-\Lambda_*) \ln\left(\frac{p_{max}+1-\Lambda_*}{1-\Lambda_*}\right),
\end{equation}
This can be further simplified for $p_{max}\gg 1$ (or $\ell\gg a$) yielding a constant (saturation) value for $\Lambda_*$ as
\begin{equation}
   \Lambda_* \rightarrow \chi p_{max} = {g\ell \over 4a K^2}\qquad d\rightarrow\infty.
   \label{eq:Lambda_infty}
\end{equation}

\begin{figure}[t]
\begin{center}
\includegraphics[width=8.cm,angle=0]{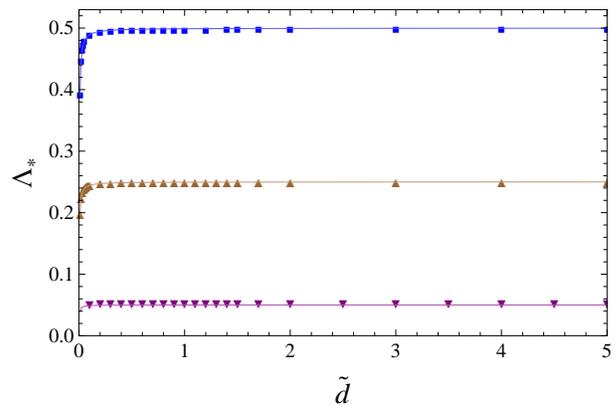}
\caption{The numerically calculated parameter $\Lambda_*$ plotted as a function of the re-scaled separation
${\tilde d}$ for $\ell/a=100$ and $g/K^2=0.02, 0.01, 0.002$ from top to bottom.
The solid curves show the  approximate expression (\ref{eq:big_lambda_2}).  
}
\label{fig2}
\end{center}
\end{figure}

For small inter-surface separations, $\tilde d \ll 1$, one can use the assumption that $ p \coth (p\tilde d)\gg |1-\Lambda_*|$ (as one can note that $\coth (p\tilde d)\simeq 1/(p\tilde d)+p\tilde d/3+\cdots$), which thus gives $\Lambda_*$ from Eq.  (\ref{eq:big_lambda}) as
\begin{equation}
\Lambda_*\simeq {\chi\over \tilde d}\,\ln (\cosh(p_{max} \tilde d)).
\label{eq:big_lambda_2}
\end{equation}
In the limit of $\tilde d\rightarrow 0$, this  gives the asymptotic behavior $\Lambda_*\simeq \chi p_{max}^2\tilde d/2$. Note that Eq. (\ref{eq:big_lambda_2}) reproduces the large $\tilde d$ behavior as well [Eq. (\ref{eq:Lambda_infty})] and therefore gives a good estimate for $\Lambda_*$ in the whole range of inter-surface separations. This is shown in Fig.~\ref{fig2}, where we plot the numerically calculated values of $\Lambda_*$ (symbols) along with the analytical estimate in Eq. (\ref{eq:big_lambda_2}) (solid lines) for several different values of the system parameters.

It is thus clear that the effects of disorder diminish  in the regime of small separations, $\tilde d\rightarrow 0$, and also when the disorder variance goes to zero, or $\chi \rightarrow 0$, since in both these limits we have $\Lambda_*\rightarrow 0$ and thus we obtain
the standard, disorder-free pseudo-Casimir interaction pressure $\Pi_0(\tilde d)$, i.e., $\Pi(\tilde d) \rightarrow \Pi_0(\tilde d)$, where
\begin{eqnarray}
&&\hspace{-.4cm}\Pi_0(\tilde d) = \!
\frac{1}{\pi}\!\!\int_{0}^{\infty}\!\!\!\frac{p^2~dp}{1+\frac{p+ 1}{p-1}\, e^{2p{\tilde d}}} \simeq \left\{
\begin{array}{ll}
     \frac{3\zeta(3)}{16\pi\tilde d^3} &\qquad \tilde d\rightarrow 0,\\
    &\\
     -\frac{\zeta(3)}{4\pi\tilde d^3} &\qquad \tilde d\rightarrow \infty.\\
    \end{array}
\right.
\label{eq:f_no_disorder}
\end{eqnarray}
These two limiting forms have been derived before (see, e.g., Refs.
\cite{rudi2,ziherl,LC1,LC2}) and show that in the disorder-free case and for $\tilde d\rightarrow 0$, the interaction pressure is repulsive and given by a universal expression
characteristic of systems with anti-symmetric boundary conditions (i.e., in this case, a LC cell with a strong homeotropic anchoring at one surface and no anchoring at the other surface). The limiting expression for $\tilde d\rightarrow \infty$, on the other hand, coincides with the universal attractive interaction due to the standard thermal Casimir effect \cite{Casimir1,Casimir2} in the case of similar (Dirichlet) boundary conditions (this is also evident because we have taken a positive mean anchoring energy per unit area $W_0>0$ at one surface and a strong homeotropic anchoring at the other).

\begin{figure}[t]
\begin{center}
\includegraphics[width=8.cm,angle=0]{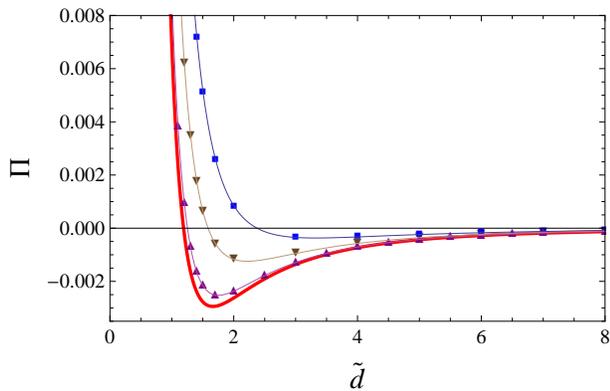}
\caption{The rescaled pseudo-Casimir pressure, $\Pi(\tilde d)$, as a function of the rescaled separation ${\tilde d}$ for $\ell/a=100$ and $g/K^2=0.02, 0.01, 0.002$ from top to bottom. The thick (red) solid curve shows the disorder-free  result for $g=0$ [Eq. (\ref{eq:f_no_disorder})].  The thin solid curves show the rescaled pressure obtained by inserting the approximate expression (\ref{eq:big_lambda_2}) into Eq. (\ref{eq:force_res0}).}
\label{fig3}
\end{center}
\end{figure}

For large separations, $\tilde d\gg 1$, the pressure in Eq. (\ref{eq:force_res0}) can be approximated by defining $u = p\tilde d$ as 
\begin{equation}
\Pi(\tilde d) \simeq \frac{1}{\pi \tilde d^3}\!\int_{0}^{\infty}\!\frac{u^2~du}{1-e^{2u}} = -\frac{\zeta(3)}{4\pi \tilde d^3}.
\end{equation}
Therefore, even though the parameter $\Lambda_*$ tends to a finite, saturated value and thus the system is  represented by a renormalized  anchoring energy, $\tilde {\mathcal W}_{eff}=1-\Lambda_*$, in the presence of the disorder (see Eqs. (\ref{eq:eff_W2}) and (\ref{eq:Lambda_infty})), the large-distance behavior of the pressure is  given by a  universal expression that coincides with that of the disorder-free, pseudo-Casimir result, Eq. (\ref{eq:f_no_disorder}).

The situation turns out to be different in the regime of intermediate inter-surface separations. The behavior of the rescaled pressure in this regime can be examined by numerically solving Eqs.  (\ref{eq:big_lambda}) and (\ref{eq:force_res0}). The results are shown in Fig.~\ref{fig3} (symbols), where we plot the rescaled pressure as a function of the rescaled inter-surface separation for several different values of the disorder variance. As seen, the effects of disorder are strongest in the regime of intermediate separations and lead to large deviations from the standard, disorder-free pseudo-Casimir interaction (thick, red solid line), which is reproduced as the limiting law for both small and large separations, in accord with our preceding asymptotic analyses. As noted before, the expression for
the pressure, Eq. (\ref{eq:force_res0}),
clearly attests to the fact that the surface disorder renormalizes the anchoring parameter (or the extrapolation length) and since the renormalized anchoring parameter, $\tilde {\mathcal W}_{eff}$, decreases monotonically  as the separation or the disorder variance are increased, one can generally expect to find a gradually more repulsive interaction force in these cases. We can thus conclude based on our numerical results that this general expectation holds only in the intermediate range of separations. We also find that in this regime the interaction pressure at a particular inter-surface separation can change sign, from attractive to repulsive, as the disorder variance is increased. As a result, the distance at which the pressure vanishes shifts to larger values, indicating that the stable bound-state between the two surfaces becomes gradually more unstable (see Fig.~\ref{fig3}).

In Fig. \ref{fig3}, we also show analytical estimates (thin solid lines) for the pressure that are obtained by inserting the approximate expression (\ref{eq:big_lambda_2}) into Eq. (\ref{eq:force_res0}). As seen, these estimates describe the behavior of the pressure to a very good extent and in the whole range of inter-surface separations.

It is also important to note that for sufficiently large disorder variances, the renormalized anchoring parameter $\tilde {\mathcal W}_{eff}$ can change sign; in other words, the homeotropic character of the disordered surface (with $W_0>0$) can change and give rise to a regime where the system may exhibit frustration resulting in a structural transition to a non-uniform ground state. In order to avoid this situation, we should have $\tilde {\mathcal W}_{eff}>0$ or $\Lambda_*<1$ that can be satisfied if the large-distance, saturation value of $\Lambda_*$ is kept below unity, i.e.
$ g\ell/(4a K^2)<1$. In the opposite case one would first need to analyze the nature of the new ground state.

\section{Conclusion}
\label{sec:con}

In this paper, we have considered the problem of fluctuation-induced or pseudo-Casimir interactions in a nematic LC film confined between two planar surfaces and, by assuming that one of the surfaces imposes an (infinitely) strong homeotropic anchoring,
we have investigated the effects of annealed disorder (randomness) in the distribution of anchoring energies on the other surface. The disordered surface exhibits a finite and positive mean anchoring energy and thus, in the absence of disorder, it also imposes a homeotropic boundary condition. In this situation 
the resulting pseudo-Casimir force in the absence of disorder is known to be  attractive at large separations and repulsive at small separations; it is given by long-ranged, universal expressions that scale as $\sim 1/d^3$ with the inter-surface distance, $d$, in both these limits.

We show that these features can change qualitatively in the presence of annealed, anchoring energy disorder.
We find that the resulting interaction force in the presence of disorder can be written in the form of a modified pseudo-Casimir force with  an effective (renormalized) anchoring energy parameter. The anchoring parameter is renormalized in such a way that, at large inter-surface separations, it saturates to a constant value that depends on the disorder variance, while, at small inter-surface separations, it tends to the mean anchoring energy, indicating that the disorder effects vanish in the limit of vanishing separation distance. However, it turns out that the interaction force tends to the disorder-free pseudo-Casimir force both at small and large inter-surface separations, where the effective extrapolation length becomes unimportant. The disorder effects are strongest in the intermediate range of separations and tend to make the inter-surface force more repulsive. Thus, also, the stable bound-state of the two surfaces (defined by the point of zero pressure) becomes continuously  more unstable as the anchoring energy variance on the disordered surface is increased.

While pseudo-Casimir interactions mediated by a nematic layer with (annealed) disordered
anchoring energy off-hand look very similar to the case of Casimir interactions between
layers with (annealed) disordered, monopolar charge distributions \cite{ali-rudi,rudiali,partial,disorder-PRL,jcp2010,pre2011,epje2012,jcp2012}, a detailed analysis shows that there are also
important differences between the two systems. Part of these differences is due to
our choice of an asymmetric system in the nematic case, which can give rise to a change
in sign of the pseudo-Casimir interaction for the same reason that there are repulsive
Casimir interactions in asymmetric layers. But a more conspicuous difference is the
behavior of the anchoring disorder that stems from a separation-dependent,  in addition to variance-dependent,  renormalization of the effective surface anchoring energy.

\begin{acknowledgments}

R.P. acknowledges ARRS grants J1-4297 and P1-0055.

\end{acknowledgments}


\begin{thebibliography}{99}

\bibitem{Casimir1}
V. A. Parsegian, {\em Van der Waals Forces} (Cambridge University Press, Cambridge, 2005).

\bibitem{Casimir2}
M. Bordag, G. L. Klimchitskaya, U. Mohideen and V. M. Mostepanenko, {\em  Advances in the Casimir Effect} (Oxford University Press, New York, 2009).

\bibitem{Mostepanenko}
V. M. Mostepanenko and N.N. Trunov, {\em The Casimir Effect and Its Applications} (Oxford University Press, 1997).

\bibitem{Kardar_rev}
M. Kardar and R. Golestanian, Rev. Mod. Phys. {\bf 71}, 1233 (1999).

\bibitem{de-gennes}
P. G. de Gennes and J. Prost, {\em The Physics of Liquid Crystals} (Oxford Science Publications, Oxford, 1995).

\bibitem{terentjev}
M. Warner and E. M. Terentjev, {\em Liquid Crystals Elastomers} (Oxford University Press, Oxford, 2003).

\bibitem{lubensky} 
P. M. Chaikin and  T. C. Lubensky, {\em Principles of Condensed Matter Physics} (Cambridge University Press,
Cambridge, 1995).

\bibitem{LC1}
A. Ajdari, L. Peliti and J. Prost, Phys. Rev. Lett. {\bf 66}, 1481 (1991).

\bibitem{LC2}
A. Ajdari, B. Duplantier, D. Hone, L. Peliti and J. Prost, J. Phys. II {\bf 2}, 487 (1992).

\bibitem{rudijure}
J. Dobnikar and R. Podgornik,  Europhys. Lett. {\bf 53}, 735 (2001).

\bibitem{rudi1}
P. Ziherl, R. Podgornik and S. \v{Z}umer, Chem. Phys. Lett. {\bf 295}, 99 (1998).

\bibitem{rudi2}
P. Ziherl, R. Podgornik and S. \v{Z}umer, Phys. Rev. Lett. {\bf 82}, 1189 (1998).

\bibitem{rudi3}
P. Ziherl, R. Podgornik and S. \v{Z}umer, Phys. Rev. Lett. {\bf 84}, 1228 (2000).

\bibitem{rudi4}
P. Ziherl, R. Podgornik and S. \v{Z}umer, J. Phys.: Condens. Matter. {\bf 12}, A221 (2000).

\bibitem{ziherl}
P. Ziherl, F. K. Haddadan, R. Podgornik and S. \v{Z}umer, Phys. Rev. E {\bf 61}, 5361 (2000).

\bibitem{haddadan}
F. K. Haddadan, D. Allender and S. \v{Z}umer, Phys. Rev. E {\bf 64}, 061701 (2001).

\bibitem{karimi}
F. Karimi Pour Haddadan and S. Dietrich, Phys. Rev. E {\bf 73}, 051708 (2006).

\bibitem{karimi2}
F. Karimi Pour Haddadan, F. Schlesener and S. Dietrich, Phys. Rev. E {\bf 70}, 041701 (2004).

\bibitem{ali-rudi}
A. Naji and R. Podgornik,  Phys. Rev. E {\bf   72}, 041402 (2005).

\bibitem{rudiali}
R. Podgornik and A. Naji, Europhys. Lett. {\bf 74}, 712 (2006).

\bibitem{partial}
Y.Sh. Mamasakhlisov, A. Naji, R. Podgornik, J. Stat. Phys. {\bf 133}, 659 (2008).

\bibitem{disorder-PRL}
A. Naji, D.S. Dean, J. Sarabadani, R. Horgan, R. Podgornik, Phys. Rev. Lett.
{\bf 104}, 060601 (2010).

\bibitem{jcp2010}
J. Sarabadani, A. Naji, D.S. Dean, R.R. Horgan and R. Podgornik, J. Chem. Phys.  {\bf 133}, 174702 (2010).

\bibitem{pre2011}
D.S. Dean, A. Naji and R. Podgornik, Phys. Rev. E {\bf 83}, 011102 (2011).

\bibitem{epje2012}
A. Naji, J. Sarabadani, D.S. Dean and R. Podgornik, Eur. Phys. J. E {\bf 35},  24 (2012).

\bibitem{jcp2012}
V. Rezvani, J. Sarabadani, A. Naji and R. Podgornik, J. Chem. Phys.  {\bf 137}, 114704 (2012).

\bibitem{dean1}
D.S. Dean, R.R. Horgan, A. Naji and R. Podgornik, Phys. Rev. E {\bf 81}, 051117 (2010).

\bibitem{dean2}
D.S. Dean, R.R. Horgan, A. Naji and R. Podgornik, Phys. Rev. A {\bf 79}, 040101(R) (2009).

\bibitem{book}
D.S. Dean, J. Dobnikar, A. Naji and R. Podgornik, {\em Electrostatics of Soft and Disordered Matter} (Pan Stanford Publishing, Singapore, 2014).

\bibitem{Review}
D. Dean. A. Naji and R. Podgornik, in preparation (2013).

\bibitem{LC3}
H. Li and M. Kardar, Phys. Rev. Lett. {\bf 67}, 3275 (1991). H. Li and M. Kardar, Phys. Rev. A {\bf 46}, 6490 (1992).

\bibitem{radzihovsky1}
Q. Zhang and L. Radzihovsky, Phys. Rev. E {\bf 81}, 051701 (2010).

\bibitem{radzihovsky2}
L. Radzihovsky and Q. Zhang, Phys. Rev. Lett. {\bf 103}, 167802 (2009).


\bibitem{edisorder}
F. Karimi Pour Haddadan, A. Naji, A. Kh. Seifi  and R. Podgornik, J. Phys: Condens matter {\bf 26}, 075103 (2014); {\em ibid} {\bf 26}, 179501 (2014).

\bibitem{LC-disorder3}
F. Karimi Pour Haddadan, A. Naji, and R. Podgornik, work in progress.



\bibitem{petridis}
L. Petridis and E. M. Terentjev, Phys. Rev. E {\bf 74}, 051707 (2006).

\bibitem{EdwardsGupta}
A.M. Gupta and S.F. Edwards, J. Chem. Phys. {\bf 98}, 1588 (1993).



\bibitem{rudi5}
R. Podgornik, Chem. Phys. Lett. {\bf 144}, 503 (1988).



\end{thebibliography}
\end{document}